\documentclass[
    4apaper
    ,final            % use final for the camera ready runs
  ]
  {aipproc}

\layoutstyle{6x9}

\newcommand{\paperone}{Paper~1}
\newcommand{\papertwo}{Paper~2}

\begin{document}

\title{The role of active galactic nuclei in galaxy formation}


\author{P.~A.~Thomas}{
  address={Astronomy Centre, University of Sussex, Falmer, Brighton, BN1 9QH, UK}
}

\author{B.~M.~Henriques}{
  address={Astronomy Centre, University of Sussex, Falmer, Brighton, BN1 9QH, UK}
  ,altaddress={Institute of Cosmology and Gravitation, University of
 Portsmouth, Portsmouth, PO1 3FX, UK}
}

\begin{abstract}
 We use Monte-Carlo Markov chain techniques to constrain
 acceptable parameter regions for the Munich L-Galaxies semi-analytic
 galaxy formation model.  Feedback from active galactic nuclei (AGN)
 is required to limit star-formation in the most massive galaxies.
 However, we show that the introduction of tidal stripping of dwarf
 galaxies as they fall into and merge with their host systems can lead
 to a reduction in the required degree of AGN feedback.  In addition,
 the new model correctly reproduces both the metallicity of large
 galaxies and the fraction of intracluster light.
\end{abstract}

\maketitle

\section{Introduction}

This paper describes the implementation of a model for dwarf galaxy
disruption within the semi-analytic framework of \citet[hereafter
  DLB07]{LuB07}.  The original model was suggested by \citet{HBT08} as
a way of both reducing the excess of dwarf galaxies and creating the
intracluster light (ICL); however this was implemented \emph{a
  posteriori}, acting only to reduce the dwarf population at the
current day.  The new model follows the stripping of dwarfs as they
fall into the halos of their parent galaxy, thus gradually reducing
their mass: affecting their infall rates, increasing the time-scale
for, and decreasing the magnitude of, the merger with the central
object.

For the purposes of these conference proceedings, the most important
result is that the masses of the black holes are reduced, with a
corresponding reduction in the level of feedback of AGN energy into
the interstellar medium.  This should be read in conjunction with the
paper by Chris Short in this volume that investigates the degree to
which the accretion energy is needed to provide feedback into the
intracluster medium (ICM) in order to provide the observed entropy
excess in clusters.

\section{Method}

Any semi-analytic model has a large number of parameters whose values
need to be optimized.  These are not all independent and correlations
between them can give insight into the key physical processes that are
constrained by the observations.
DLB07 has 12 explicit parameters (as well as lots of hidden ones).  We
fix all but 6 parameters in our analysis:
\begin{itemize}
\item the star formation efficiency, $\alpha_{\rm SF}$
\item the AGN radio mode efficiency, $k_{\rm AGN}$
\item the black hole growth efficiency, $f_{\rm{BH}}$
\item the supernova reheating and ejection efficiency,
respectively $\epsilon_{\rm{disk}}$ and $\epsilon_{\rm{halo}}$
\item the ejected gas reincorporation efficiency, $\gamma_{\rm ej}$
\end{itemize}

In \citet[hereafter \paperone]{HTO09} we introduced a Monte-Carlo
Markov Chain (MCMC) method for mapping out allowable likelihood
regions in parameter space and used it to find the best-fitting
values for the above parameters, using as constraints the $K$-band
luminosity function, the $B-V$ colours, and the black hole-bulge mass
ratio.   Subsequently we have introduced a self-consistent stripping
model for the dwarfs \citep[hereafter \papertwo]{HeT09}.  We summarise
some of the results from that paper below.

We build our model on merger trees derived from the \emph{Millennium
Simulation} \citep{SWJ05}, using a small but representative subset of
the total volume for our MCMC analysis.  However, the results
presented below come from applying the best-fit model to the entire
simulation.

\section{Results}

The best fit and confidence limits for the 6 free parameters in the
model with satellite disruption, together with the published values
from DLB07 and the best fit for the model without satellite disruption
are shown in Table \ref{tab:dismargestats}.  The most significant
thing to note is that, in the disruption model, the star-formation
rate is much enhanced and the rate of re-incorporation of expelled gas
is reduced.  This results in a much lower fraction of cold gas in the
interstellar medium and consequently a lower accretion rate onto the
central black hole.  Thus the black hole masses are smaller in the
disruption model, in better agreement with observations, and
the feedback rates from AGN are reduced.

\begin{table}
\begin{tabular}{lccccccc}
  \\[3pt]
  \hline
  & \tablehead{1}{c}{t}{DLB07} 
  & \tablehead{1}{c}{t}{\paperone} 
  & \tablehead{1}{c}{t}{\papertwo} 
  & $\boldmath{-2\sigma}$  
  & $-1\sigma$  
  &+1$\sigma$
  &+2$\sigma$\\
  \hline
  $\alpha_{\rm{SF}}$ &0.03 &0.039 &0.17 &0.078 &0.13 &0.28 &0.53\\
  $k_{\rm{AGN}}$ &$7.5\times10^{-6}$ &$5.0\times10^{-6}$& $5.3\times10^{-6}$ &$2.7\times10^{-6}$ &$3.7\times10^{-6}$ &$6.2\times10^{-6}$&$7.9\times10^{-6}$\\
  $f_{\rm{BH}}$ &0.03 &0.032 &0.047 &0.030 &0.041 &0.061 &0.075 \\ 
  $\epsilon_{\rm{disk}}$ &3.5 &10.28 &6.86 &5.22 &6.33 &8.51 &10.11 \\
  $\epsilon_{\rm{halo}}$ &0.35 &0.53 &0.33 &0.26 &0.31 &0.40 &0.46\\
  $\gamma_{\rm{ej}}$ &0.5 &0.42 &0.13 &0.076 &0.12 &0.24 &0.30 \\
  \hline
\end{tabular}
%}
  \caption{Statistics from the MCMC parameter estimation for the
    parameters in the satellite disruption model (\papertwo). The best
    fit and marginalized confidence limits are compared with the
    published values from DLB07 and with the best fit values obtained
    without the inclusion of satellite disruption
    (\paperone).}  \label{tab:dismargestats}
\end{table}

%% \begin{figure}
%% \centering
%%   \includegraphics[width=.8\textwidth]{dismcmch2_bestfit_icl}
%%   \caption{Fraction of the mass in the ICM over the
%%   total stellar mass of the group as a function of virial mass. The
%%   blue dots are a representative sample of the total galaxy population
%%   in the best fit model with satellite disruption. The solid and
%%   dashed lines represent the median of the $M_{\rm
%%     {ICL}}$/$M_{\rm{total}}$ distribution for the satellite disruption
%%   model with the best fit and original parameters respectively.}
%%   \label{fig:disbestfiticl}
%% \end{figure}

\begin{figure}
 \centering
 \includegraphics[width=.8\textwidth]{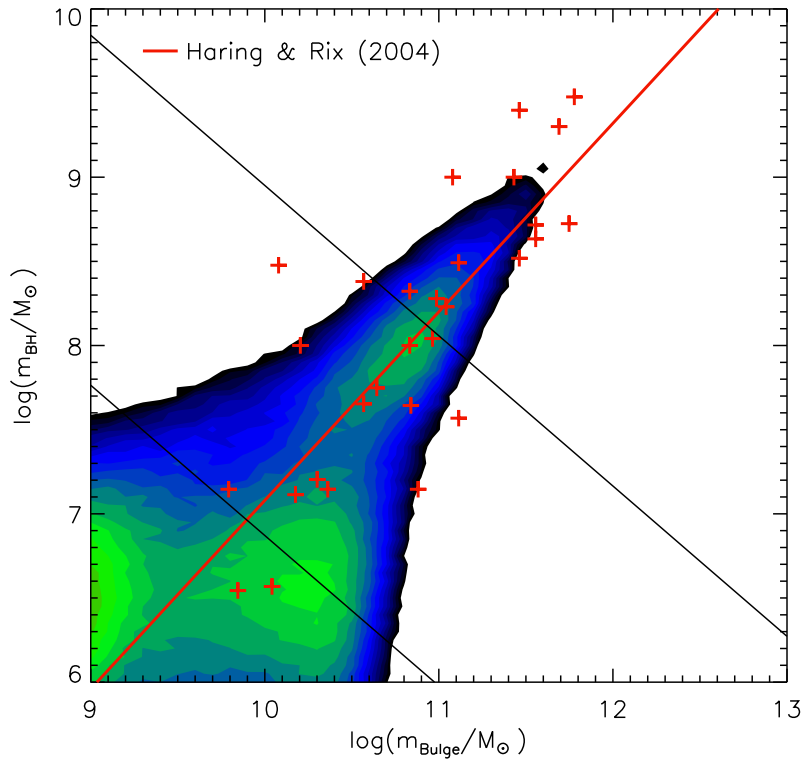}
 \caption{The black hole-bulge mass relation for the satellite
   disruption model (solid contours). The red crosses represent
   observations from \citet{HaR04} with the best fit to the data
   points given by the red line.}
\label{fig:disbhbm}
\end{figure}

Figure~\ref{fig:disbhbm} shows a comparison of the model
black-hole/bulge mass relation (contours) with observations (crosses).
Although the observations are biased to high-mass systems, the
distribution of systems above and below the best-fit line are correct. 

\begin{figure}
  \includegraphics[width=.8\textwidth]{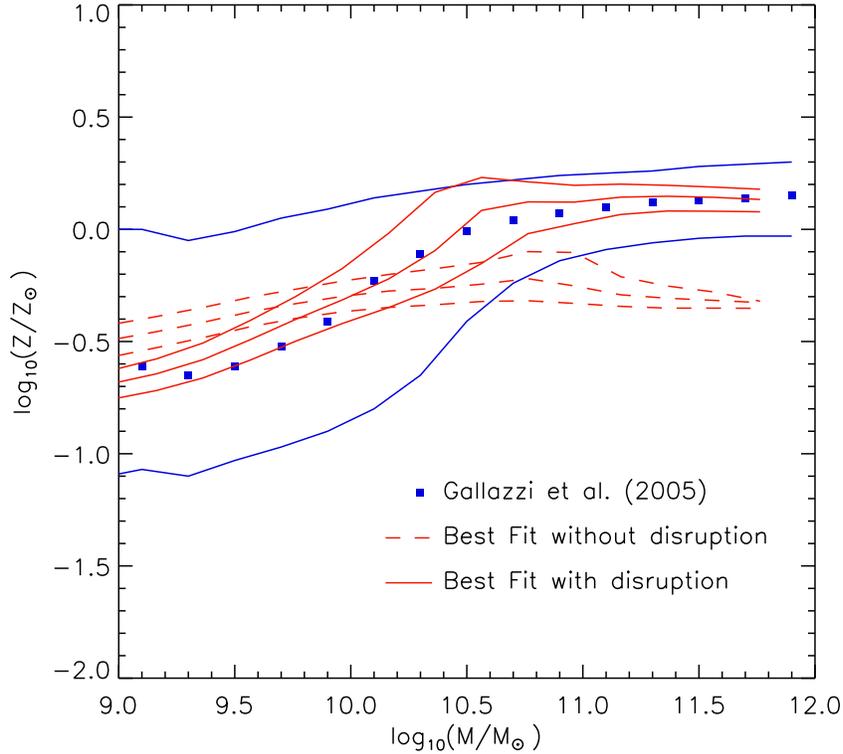}
  \caption{Comparison between the metallicity stars in the best fit for
  the satellite disruption model (solid red lines), in the best fit
  for DLB07 (dashed red lines) and in observations from
  \citet{GCB05} (blue squares and lines). For all the data
  sets, the central line represents the median value of metallicity in
  each mass bin (the blue squares for the observational data), while
  the upper and lower lines represent the 16th and 86th percentiles of
  the distribution.}
  \label{fig:disbestfitmetals}
\end{figure}

Although not used to constrain the model, it is gratifying that it
produces the correct amount of ICL (see Figure~9 of \papertwo).
In addition, as shown in Figure~\ref{fig:disbestfitmetals}, it
provides a much better match to the metallicity distribution of
massive galaxies than the original model (in which massive galaxies
accrete too many low-metallicy stars during mergers of dwarf satellites.)

\section{Discussion}

The black hole growth model presented here provides a real challenge
for models that require AGN heating to raise the entropy of the
ICM. Even using the original DLB07 model, with its higher black hole
masses, 35 per cent of the available rest mass energy is required (see
article by Chris Short in this volume, also \citet{ShT09}). Similar
conclusions have been reached by \citet{BMB08} using the Durham
GALFORM semi-analytic model.

\begin{theacknowledgments}

This work was undertaken using the Virgo Consortium cluster of
computers, COSMA.

BH acknowledges the support of his PhD scholarship from the Portuguese
Science and Technology Foundation which supported him for most of the
time while this work was developed.  PAT was supported by an STFC
rolling grant.
\end{theacknowledgments}

\bibliographystyle{aipproc}
\bibliography{thomas_p}

\end{document}